\documentclass[letter]{aa} 
\usepackage{graphicx}
\usepackage{txfonts}
\usepackage{natbib}
\usepackage[breaklinks=false]{hyperref}
\newcommand\dif{{\ensuremath{\mathrm{d}}}}
\newcommand\dt{{\ensuremath{\dif t}}}
\newcommand\dr{{\ensuremath{\dif r}}}
\newcommand\dth{{\ensuremath{\dif\theta}}}
\newcommand\dphi{{\ensuremath{\dif\varphi}}}
\newcommand\ds{{\ensuremath{\dif s}}}
\newcommand\Ms{{\ensuremath{\mathrm{M}_{\odot}}}}
\newcommand{\Mpy}{\Ms\,{\rm yr}{\ensuremath{^{-1}}}}
\newcommand{\dm}{\ensuremath{\dot M}}
\newcommand{\gva}{{\sc genec}}

\usepackage{color}

\begin{document}

\title{Rapidly accreting supermassive stars: reliable determination of the final mass}

\author{L. Haemmerl\'e}
\authorrunning{Haemmerl\'e}

\institute{D\'epartement d'Astronomie, Universit\'e de Gen\`eve, chemin des Maillettes 51, CH-1290 Versoix, Switzerland}


 
\abstract
{Supermassive black holes might form by direct collapse, with a supermassive star (SMS) as progenitor.
In this scenario, the SMS accretes at $>0.1$ \Mpy\ until it collapses into a massive black hole seed due to the general-relativistic (GR) instability.
However, the exact mass at which the collapse occurs is not known, as existing numerical simulations give divergent results.}
{Here, this problem is addressed analytically, which allows for {\it ab initio}, reliable determination of the onset point of the GR instability,
for given hydrostatic structures.}
{We apply the relativistic equation of radial pulsations in its general form to the hydrostatic \gva\ models already published.}
{We show that the mass of spherical SMSs forming in atomically cooled haloes cannot exceed 500\,000 \Ms, in contrast to previous claims.
On the other hand, masses in excess of this limit, up to $\sim10^6$ \Ms, could be reached in alternative versions of direct collapse.}
{Our method can be used to test the consistency of GR hydrodynamical stellar evolution codes.
}
 
 
\maketitle
%

\section{Introduction}
\label{sec-in}

Supermassive stars (SMSs) are currently the main candidates for being the progenitors of the most distant supermassive black holes
(e.g.~\citealt{woods2019,haemmerle2020}).
Two formation channels have been proposed in the literature:
(1) atomically cooled haloes, where primordial atomic gas inflows at 0.1 -- 10 \Mpy\ with negligible fragmentation
(e.g.~\citealt{bromm2003b,latif2013e,regan2017});
(2) major mergers of gas-rich galaxies \citep{mayer2010,mayer2015,mayer2019},
where inflows as high as $10^5$ \Mpy\ can be reached, for metallicities up to solar.

SMSs accreting at rates $>0.1$ \Mpy\ have been found to evolve as red supergiant protostars with an extended radiative envelope
\citep{hosokawa2013,sakurai2015,haemmerle2018a,haemmerle2019c},
and to collapse via the general-relativistic (GR) instability \citep{chandrasekhar1964} at masses $10^5-10^6$ \Ms\ \citep{umeda2016,woods2017}.
However, beyond orders of magnitude, previous attempts to determine the exact final mass of SMSs failed to converge,
as unexplained discrepancies remain in the outcome of the various hydrodynamical codes \citep{woods2019}.

In the present work, we show that the use of numerical hydrodynamics is not the relevant method to capture the onset of GR instability.
Instead, we propose an analytical {\it ab initio} treatment of hydrodynamics,
which consists in the simple application of the pulsation equation derived by \cite{chandrasekhar1964} in its general form.
Provided suitable transformations that preserve its generality,
this equation allows for reliable and accurate determination of the mass at which the GR instability develops.
We apply this method to the hydrostatic models published in \cite{haemmerle2018a,haemmerle2019c} and estimate the final masses.

\section{Method}
\label{sec-meth}

The relativistic equation of stellar pulsations has been derived from Einstein's equations by \cite{chandrasekhar1964}
under the assumptions of spherical symmetry and adiabatic perturbations to dynamical equilibrium.
For a simple choice of the trial function that characterises the displacement from equilibrium,
the equation for the frequency $\omega$ of global pulsations can be written as
\begin{eqnarray}
{\omega^2\over c^2}I_0=\sum_{i=1}^4I_i
\label{eq-chandra}\end{eqnarray}
with
\begin{eqnarray}
I_0&=&\int_0^Re^{a+3b}(P+\rho c^2)r^4\dr					\label{eq-I0}\\
I_1&=&9\int_0^R\Gamma_1e^{3a+b}Pr^2\dr					\label{eq-I1}\\
I_2&=&4\int_0^Re^{3a+b}P'r^3\dr							\label{eq-I2}\\
I_3&=&{8\pi G\over c^4}\int_0^Re^{3(a+b)}P(P+\rho c^2)r^4\dr		\label{eq-I3}\\
I_4&=&-\int_0^Re^{3a+b}{P'^2\over P+\rho c^2}r^4\dr			\label{eq-I4}
\end{eqnarray}
where $P$ is the pressure, $\rho$ the density of relativistic mass,
$r$ the radial coordinate ($R$ the radius of the star), $c$ the speed of light, $G$ the gravitational constant, $\Gamma_1$ the first adiabatic exponent,
and $a$ and $b$ the coefficients of the metric:
\begin{equation}
\ds^2=-e^{2a}(c\dt)^2+e^{2b}\dr^2+r^2\dth^2+r^2\sin^2\theta\ \dphi^2
\label{eq-ds}\end{equation}
The symbol $'$ indicates the derivatives with respect to $r$.
The metric can be derived from the other quantities through the well known expressions for spherical hydrostatic configurations:
\begin{eqnarray}
a'&=&{GM_r\over r^2c^2}{1+{4\pi r^3\over M_rc^2}P\over1-{2GM_r\over rc^2}}	,\qquad
e^{2a(R)}=1-{2GM\over Rc^2}\\
e^{-2b}&=&1-{2GM_r\over rc^2}
\label{eq-metric}\end{eqnarray}
where $M_r$ is the relativistic mass-coordinates and $M=M_R$.
The GR instability develops when the right-hand side of equation (\ref{eq-chandra}),
i.e. the sum of integrals (\ref{eq-I1}-\ref{eq-I2}-\ref{eq-I3}-\ref{eq-I4}), becomes negative, since $I_0$ is always positive.
It occurs when the absolute value of $I_2+I_4$ exceeds that of $I_1+I_3$,
because $I_2<0$ due to the pressure gradient, and obviously $I_4<0$ while $I_1$ and $I_3$ are positive.

In principle, equation~(\ref{eq-chandra}) could be applied directly to any hydrostatic model accounting for GR corrections.
However, in practice, this method hardly captures the GR instability in numerical models, as it will be illustrated below.
On the other hand, we will show that a simple integration by parts of $I_2$ allows to solve this problem.
We write
\begin{eqnarray}
4\int_0^Re^{3a+b}P'r^3\dr=
&&-12\int_0^Re^{3a+b}\left(a'+{b'\over3}\right)Pr^3\dr	\nonumber\\
&&-12\int_0^Re^{3a+b}Pr^2\dr
\end{eqnarray}
(boundary terms vanish).
With this integration by parts, equation (\ref{eq-chandra}) can be rewritten as
\begin{eqnarray}
{\omega^2\over c^2}I_0=\sum_{i=1}^4J_i
\label{eq-chandra2}\end{eqnarray}
with
\begin{eqnarray}
J_1&=&9\int_0^R\left(\Gamma_1-{4\over3}\right)e^{3a+b}Pr^2\dr	\label{eq-J1}\\
J_2&=&-12\int_0^Re^{3a+b}\left(a'+{b'\over3}\right)Pr^3\dr		\label{eq-J2}\\
J_3&=&I_3											\label{eq-J3}\\
J_4&=&I_4											\label{eq-J4}
\end{eqnarray}

The difficulties in capturing the GR instability on numerical models with equation (\ref{eq-chandra}) arises from the fact that
the GR corrections remain always small in SMSs (${2GM_r\over rc^2}\sim{P\over\rho c^2}\lesssim0.01$),
and are easily hidden by numerical noise.
In the Newtonian limit, integrals (\ref{eq-I1}-\ref{eq-I2}-\ref{eq-I3}-\ref{eq-I4}) reduce to:
\begin{eqnarray}
I_1&\to&9\int_0^R\Gamma_1Pr^2\dr>0				\label{eq-I1postNewton}\\
I_2&\to&-4\int_0^R\rho GM_rr\dr<0							\label{eq-I2postNewton}\\
I_3&\to&8\pi G\int_0^R\left({P\over\rho c^2}\right)\rho^2 r^4\dr\to0		\label{eq-I3postNewton}\\
I_4&\to&-\int_0^R\left({GM_r\over rc^2}\right)\rho GM_rr\dr\to0			\label{eq-I4postNewton}
\end{eqnarray}
where we used the classical equation of hydrostatic equilibrium in $I_2$ and $I_4$.
One can see that in the Eddington limit ($\Gamma_1\to4/3$) the integrals do not vanish individually.
Of course, their sum must vanish, since the Eddington limit corresponds to the classical limit for stability.
But this can only be guaranteed to a limited extent when the integration is made over numerical models,
which cannot satisfy the equations of structure with infinite precision.
In contrast, integrals $J_2$, $J_3$ and $J_4$ of equation (\ref{eq-chandra2}) all vanish individually in the Newtonian limit, while $J_1$ reduces to
\begin{eqnarray}
J_1\to9\int_0^R\left(\Gamma_1-{4\over3}\right)Pr^2\dr	\label{eq-J1postNewton}
\end{eqnarray}
We see that the use of equation (\ref{eq-chandra2}) makes explicit the classical limit for stability ($\Gamma_1\to4/3$),
as the Eddington and Newtonian components have been cancelled analytically.

In the present work, we apply equation (\ref{eq-chandra2}) to the stellar models published in \cite{haemmerle2018a,haemmerle2019c}.
These models were computed with \gva, accounting for GR effects through the post-Newtonian Tolman-Oppenheimer-Volkoff equation.
Accretion is included, at constant rates $\dm=0.1$ -- 1 -- 10 -- 100 -- 1000 \Mpy\ for zero metallicity and 1 -- 1000 \Mpy\ for solar metallicity.
This method allows to derive from first principles the mass at which the GR instability develops,
and thus presents a higher consistency than numerical hydrodynamics.

\section{Results}
\label{sec-res}

\begin{figure}\begin{center}
\includegraphics[width=.49\textwidth]{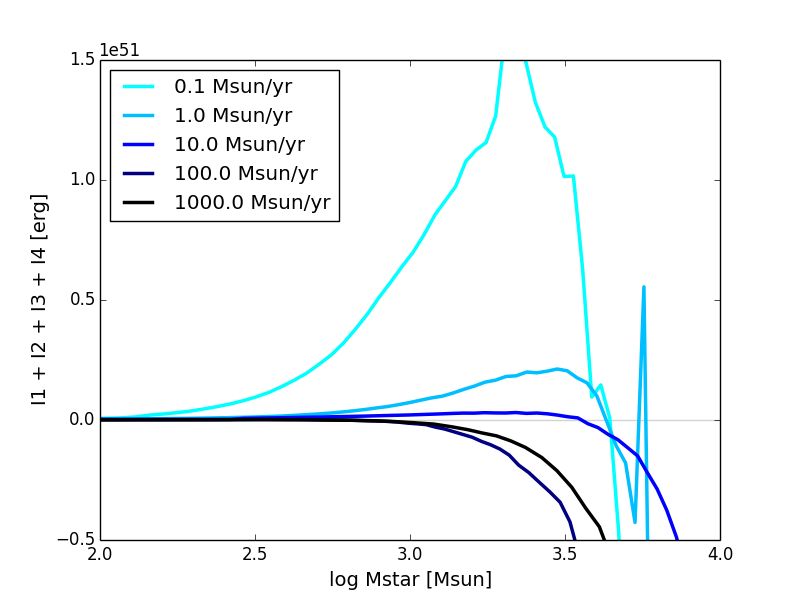}
\caption{Right-hand side of equation (\ref{eq-chandra}) evaluated on the \gva\ models at zero metallicity and indicated accretion rates.
The grey line indicates the zero.}
\label{fig-sum1}\end{center}\end{figure}

\begin{figure}\begin{center}
\includegraphics[width=.49\textwidth]{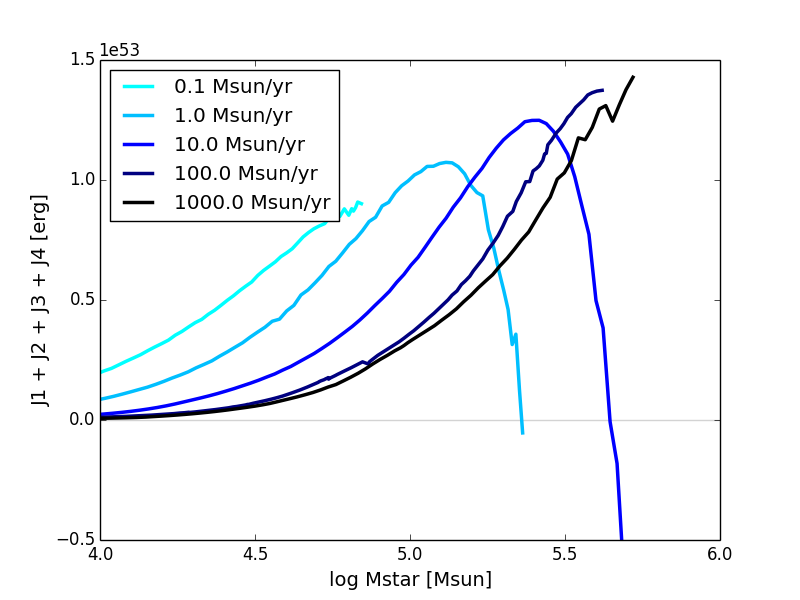}
\caption{Right-hand side of equations (\ref{eq-chandra2}) for the same models as in figure \ref{fig-sum1}.}
\label{fig-sum2}\end{center}\end{figure}

In order to illustrate the difficulty to capture the GR instability with equation (\ref{eq-chandra}),
we evaluate the right-hand side of this equation by numerical integration over the \gva\ models (figure \ref{fig-sum1}).
For all the models, the curve decreases below zero before the stellar mass reaches $10^4$ \Ms.
Figure \ref{fig-sum2} shows the same experiment with equation (\ref{eq-chandra2}), and we see a very different result.
In this case, only the models at 1 and 10 \Mpy\ reached the instability, at masses 229 000 \Ms\ and 437 000 \Ms, respectively.

\begin{figure}\begin{center}
\includegraphics[width=.49\textwidth]{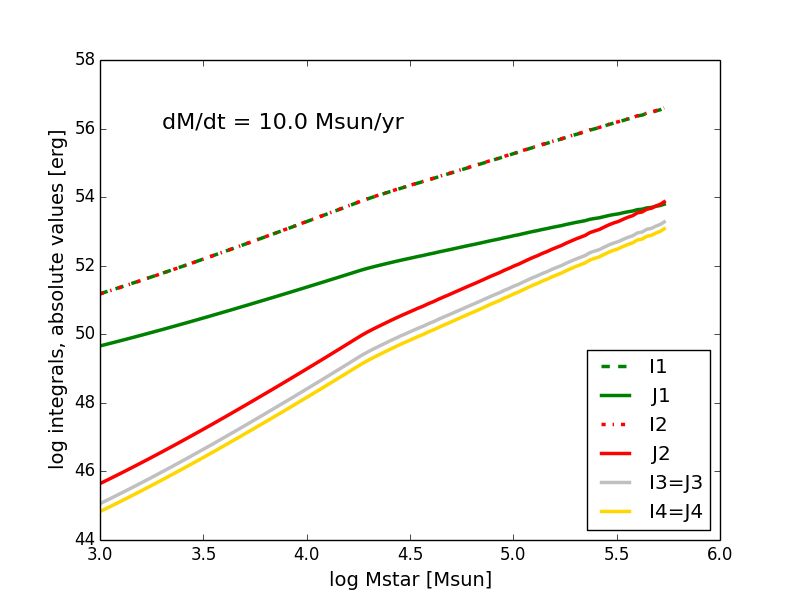}
\caption{
Absolute values of the integrals (\ref{eq-I1}-\ref{eq-I2}, dashed lines) and (\ref{eq-J1}-\ref{eq-J2}-\ref{eq-J3}-\ref{eq-J4}, solid lines)
for the \gva\ model at zero metallicity with the indicated rate.
The green curves are the stabilising integrals $I_1,J_1>0$, while the red curves are the destabilising integrals $I_2,J_2<0$.
The grey curves are $I_3>0$ and $I_4<0$, that remain negligible, and unchanged between equations (\ref{eq-chandra}) and (\ref{eq-chandra2}).}
\label{fig-int}\end{center}\end{figure}

The reliability of these two contradictory results can be tested by looking at the terms of the sum individually, i.e. the integrals $I_i$ and $J_i$.
Figure \ref{fig-int} shows these integrals in absolute values for the \gva\ model at zero metallicity and $\dm=10$~\Mpy.
The knee in the integrals near $\log M/\Ms=4.3$ indicates the beginning of H-burning.
Integrals $I_3=J_3$ and $I_4=J_4$ remain always negligible compared to the others,
and their role in the GR instability is still weakened by the fact that they have opposite signs and similar magnitudes.
The stability of the star depends mainly on the balance between the stabilising integral $I_1>0$ (or $J_1>0$), that reflects the departures from the Eddington limit,
and the destabilising integral $I_2<0$ (or $J_2<0$), that reflects the departures from the Newtonian limit.
We see that $I_1$ and $I_2$ are indeed very close, as expected for post-Newtonian stars near the Eddington limit.
But the fact that the differences remain invisible in a logarithmic scale indicates that the {\it relative} differences are negligible.
Thus, numerical noise that would not affect the equilibrium properties could hinder dynamical effects.
It shows how difficult it is to capture the GR instability simply by accounting for GR hydrodynamics on numerical models,
even if these models satisfy the equations of structures with a high precision.
In contrast, if we compare $J_1$ and $J_2$ we see that they differ by orders of magnitude during most of the run.
The point at which they cross each other is clearly recognisable, and do not suffer from the small numerical variations in the integrals,
which shows the high accuracy of this method.

The final masses obtained in this way with the \gva\ models are summarized in Table \ref{tab-mfin}.
When the model didn't reach the instability at the end of the run, we indicate the last mass reached as a lower limit.
The model at 0.1 \Mpy\ and zero metallicity is not shown, its mass at the end of the run is 70\,000~\Ms, which is far from instability (figure \ref{fig-sum2}).
We confirm that increasing the accretion rate allows to reach higher masses before the instability.
But we see that for $\dm\leq10$ \Mpy\ and $Z=0$, the stellar mass cannot exceed 500\,000 \Ms.
Masses above this limit require rates $\gtrsim100-1000$ \Mpy, or the presence of metals.
The most massive stable model we obtain is that at solar metallicity and 1000~\Mpy, where the GR instability develops only at 759 000 \Ms.
Interestingly, an extrapolation of the curve of the 1000 \Mpy\ model at zero metallicity on Figure \ref{fig-sum2}
suggests a final mass $>10^6$ \Ms.

\begin{table}
\caption{Stellar mass (in $10^5$ \Ms) at the onset of GR instability for the \gva\ models at the indicated metallicities and accretion rates.}
\label{tab-mfin}
\centering
\begin{tabular}{c|cccc|}
			& 1 \Mpy		& 10 \Mpy	& 100 \Mpy	& 1000 \Mpy	\\\hline
$Z=0$		& 2.29		& 4.37	&$(>4.25)$	&$(>5.25)$	\\
$Z=Z_\odot$	&$(>3.02)$	&		&			&7.59   		\\\hline
\end{tabular}
\end{table}

\section{Discussion}
\label{sec-dis}

\begin{figure}\begin{center}
\includegraphics[width=.49\textwidth]{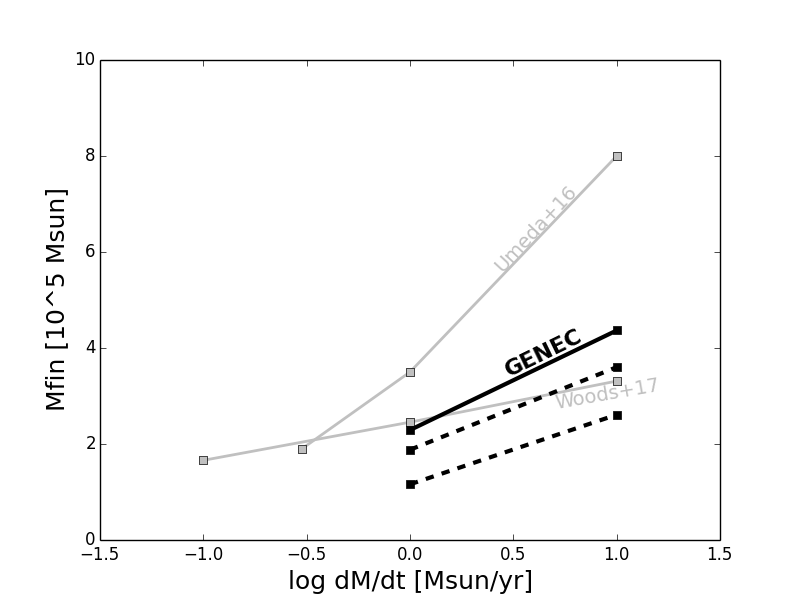}
\caption{Final masses of SMSs at zero metallicity, accreting at the rates of atomically cooled haloes.
The results of the present work are shown by a solid black line.
The masses obtained with the polytropic criterion are shown as dashed lines, when applied in the whole star (lower line) and in the convective core only (upper line).
The results of \cite{umeda2016} and \cite{woods2017}, obtained with numerical hydrodynamics, are shown in grey.}
\label{fig-mfin}\end{center}\end{figure}

Two studies addressed the question of the final mass of Population III SMSs accreting at the rates of atomically cooled haloes \citep{umeda2016,woods2017}.
These works were based on stellar evolution models accounting numerically for hydrodymamics,
with the assumptions of spherical symmetry and post-Newtonian corrections.
The only ingredient that accounts for GR in these studies is the correction to the equation of radial momentum, following \cite{fuller1986}.
With this method, the GR hydrodynamics is treated only numerically, by adding small corrections to the intrinsically stable Newtonian structures.
In spite of identical physical ingredients, these two studies obtained very different results, in particular for rates $>1$~\Mpy,
and the origin of these discrepancies is not understood \citep{woods2019}.
The final masses obtained in these simulations are shown in Figure \ref{fig-mfin} as a function of the accretion rate.
As discussed in \cite{woods2019}, the final mass of the 10~\Mpy\ model differs by a factor $>2$ between both studies.
It is thought that at least part of this discrepancy is the result of the differences that appear in the hydrostatic structures themselves,
since \cite{woods2017} obtains much larger convective cores than \cite{umeda2016}.
But this statement is speculative, and moreover the origin of the differences in the structure are not better understood.

The \gva\ models of \cite{haemmerle2018a,haemmerle2019c} used in the present work have been computed with exactly the same ingredients
as \cite{umeda2016} and \cite{woods2017}, except the hydrodynamics which is missing.
Regarding the structures, these models are in very good agreement with \cite{umeda2016},
as well as with \cite{hosokawa2013} who consider lower masses.
It allows for a direct comparison between our results, that account for hydrodynamics {\it ab initio},
and those of \cite{umeda2016} that account for hydrodynamics with numerical methods.
To that aim, we plotted in Figure \ref{fig-mfin} the final masses of the \gva\ models at zero metallicity and $\dm=1-10$ \Mpy.
We see that, in spite of the fact that our structures are identical to those of \cite{umeda2016}, our final masses remain much closer to those of \cite{woods2017}.
For 10 \Mpy, our final mass exceeds that of \cite{woods2017} by $\sim30\%$ only, while a factor 2 remains in the final mass of \cite{umeda2016}.
It reveals clearly that the differences in the size of the convective core between \cite{umeda2016} and \cite{woods2017}
count for negligible part in the discrepancies on the final mass,
and that these discrepancies rely essentially on the numerical treatment of hydrodynamics in \cite{umeda2016}.
It is visible on the entropy profiles of \cite{umeda2016} that numerical noise is significant in their model, to a larger extent than in \gva\ models.
It does not seem to have serious consequences on the hydrostatic structures themselves,
but in view of the difficulties to capture small post-Newtonian corrections in numerical models (Section \ref{sec-res}),
numerical noise in \cite{umeda2016}'s structures appears as the most natural suspect for the factor 2 discrepancies in the final mass.
Our interpretation can be verified by these authors by a simple application of our {\it ab initio} method.

We also indicate in Figure \ref{fig-mfin} the final masses obtained with the polytropic criterion of \cite{chandrasekhar1964} applied to the \gva\ models,
in the star as a whole (lower dashed line) and in the convective core only (upper dashed line).
It confirms that the use of the polytropic criterion leads to an underestimation of the final mass, as already pointed out by \cite{woods2017}.
Thus, the present method appears as the most consistant, accurate and straightforward to apply.

\section{Summary and conclusions}
\label{sec-out}

We showed how the GR instability can be captured {\it ab initio} with high accuracy in numerical hydrostatic stellar models,
which allows for reliable determination of the final mass of accreting SMSs,
in a simpler and more consistant way than numerical hydrodynamics.
We emphasise that our treatment of the instability relies directly on the general expression of the relativistic pulsation equation of \cite{chandrasekhar1964},
without any additional assumption.
According to our results, the mass of spherical Population III SMSs accreting at the rates of atomically cooled haloes remains always $<500\,000$ \Ms.
Larger masses can be reached only for rates $\gtrsim100-1000$ \Mpy, or for non-primordial chemical composition.
For conditions of galaxy mergers \citep{mayer2019,haemmerle2019c}, masses $\sim10^6$ \Ms\ could be reached.
Our method can be used to check the consistency of the numerical treatment of GR hydrodynamics in stellar evolution codes.

\begin{acknowledgements}
LH has received funding from the European Research Council (ERC) under the European Union's Horizon 2020 research and innovation programme
(grant agreement No 833925, project STAREX).
\end{acknowledgements}

\bibliographystyle{aa}
\bibliography{bib}

\end{document}